\documentclass[preprint,secnumarabic,showpacs,amssymb, nobibnotes ,aps, prd]{revtex4-1}

\usepackage{graphicx}\usepackage{dcolumn}\usepackage{bm}
\begin{document}
\title{Exact plane gravitational waves in the de Rham-Gabadadze-Tolley model of massive gravity}
\author{Morteza Mohseni}
\email{m-mohseni@pnu.ac.ir}
\affiliation{Physics Department, Payame Noor University, 19395-4697 Tehran, Iran}
\date{\today}
\begin{abstract}
We show that the nonlinear massive gravity model of de Rham, Gabadadze, and Tolley admits exact plane gravitational wave solution whose waveform obeys the two-dimensional Helmholtz 
equation. The solution is valid for arbitrary values of the coefficients of the cubic and quartic terms. In the massless limit the solution reduces to the Aichelburg-Sexl metric in general relativity.   
\end{abstract}
\maketitle
\section{Introduction}
Massive gravity theories have been of interest for the past decades. Various reasons for this interest range from pure field theoretic considerations to the recently raised problem of the 
accelerated expansion of the universe.  In fact it has been recently suggested, e.g. in Refs. \cite{chery,chery2}, that massive gravitons may be considered as an alternative to dark energy, 
explaining the accelerated expansion.  

The pioneering work by Fierz and Pauli \cite{fierz} resulted to a now well-known model of massive gravity. That model with its unique property of being the sole massive extension of the 
general theory of relativity which is linear, is faced with subtle theoretical problems such as the extra helicity degrees of freedom and the so-called van Dam-Veltman-Zakharov discontinuity \cite{vain,vain2} which 
implies that the theory does not lead to the general relativity in the zero-mass limit.  It has been attempted to remedy this by adding nonlinear terms to the Fierz-Pauli model of massive gravity 
\cite{vain1}.  However they failed to do so as shown by the analysis of Boulware and Deser  \cite{deser} according to which, an unwanted peculiarity is always associated with these models: the 
appearance of ghosts and its consequent instability.  

A great deal of progress has been recently made in this field by the new massive gravity (NMG) model proposed by Bergshoeff et.al. \cite{bergshoeff} which have proved to be unitary and 
ghost free \cite{deser2,deser1}, and there are claims that it is renormalizable \cite{deser3}. However, the latter model is restricted to three space-time dimensions.  A recent review of the 
theoretical aspects of various massive gravity models may be found in Ref. \cite{hint}, see also Ref. \cite{vali}. 

Recent works on massive gravity in the context of effective field theories \cite{add,add2} have resulted in the hope that there might be some way out of this dilemma. In a recent paper by de 
Rham et. al. \cite{rahm} a four-dimensional nonlinear massive gravity model (the dRGT model) was proposed which is free from the Boulware-Deser ghost up to and including quartic order 
away from the decoupling limit.  The dRGT model is essentially constructed  by algebraic resummation of the nonlinear terms in a previously proposed effective field theory of massive gravity 
\cite{arkani,arkani1,arkani3}. Some related works may be found in Refs. \cite{gabad1,gabad2,gabad3,lasha}. The model was used in Ref. \cite{gabad} to explain the cosmic acceleration from 0-
helicity gravitons. The Vainstein mechanism of the model was investigated in Ref. \cite{gabad6}. The effect of massive gravitons on the stellar gravitational field was studied in Ref. 
\cite{gabad5}. Strong coupling effects was studied in Ref. \cite{gabad7}. Exact Schwarzschild-de Sitter black hole solutions of  the dRGT model have been presented in Ref. \cite{new} and they 
have shown there to go smoothly to the Schwarzschild and Reissner-Nordstr\"{o}m solutions of general relativity in the zero-mass limit for a specific choice of the parameters. 

In the present work we seek exact plane gravitational wave solutions in the context of the dRGT model. This is well motivated by the current interest in gravitational waves in general and in exact 
plane wave solutions, in particular. For example, it is believed that certain classes of plane gravitational waves, namely gravitational shock waves, have dominant role in scattering 
process at ultrahigh-energy regime \cite{hooft,hooft2}. Also, the propagation of gravitational waves from various sources have been studied in order to place an upper bound on graviton 
mass, see e.g. Ref. \cite{fin} and references therein. 

In the next sections, we shall start with a brief review of the dRGT model and by solving the equations of motion, we show that the model admits exact plane gravitational waves.      
 \section{The dRGT model}
Let us represent the Minkowski metric $\gamma_{\mu\nu}$ in a chart $\{x^a\},\, a=0,1,2,3$ by
\begin{equation}\label{1a}
\gamma_{\mu\nu}=\eta_{ab}\partial_\mu\phi^a\partial_\nu\phi^b
\end{equation}
where $\eta_{ab}=diag(-1,1,1,1)$, and $\phi^a=x^a$ are the St\"{u}ckelberg fields in the unitary gauge. The space-time metric $g_{\mu\nu}$ is then incorporated via
\begin{equation}\label{1b}
{\gamma^\mu}_\nu=g^{\mu\lambda}\gamma_{\lambda\nu}.
\end{equation}
The dRGT model of massive gravity is described by the following Lagrangian density \cite{rahm}
\begin{equation}\label{1c}
{\mathcal L}=\frac{1}{2}M^2_p{\sqrt{-g}}(R-m^2U)
\end{equation}
where the first term in the parentheses is the curvature scalar which together with the coefficients outside constitute the Einstein-Hilbert Lagrangian density. The second term defines the mass
term and is given by   
\begin{equation}\label{1d}
U=a_2{\mathcal L}^{(2)}+a_3{\mathcal L}^{(3)}+a_4{\mathcal L}^{(4)}
\end{equation}
with $a_2, a_3$,  and $a_4$ being dimensionless coupling constants. The sign of $a_2$ is negative, otherwise tachyons are allowed in the model. In fact, one may  
set $a_2=-1$ without loss of generality. However, in what follows we also consider the effect of $a_2$ having the (wrong) positive sign. 

The explicit form of ${\mathcal L}^{(2,3,4)}$ can be expressed in terms of the tensor ${\kappa^\mu}_\nu$ defined by
\begin{equation}\label{1e}
{\kappa^\mu}_\nu=\delta^\mu_\nu-{\beta^\mu}_\nu
\end{equation}
where ${\beta^\mu}_\nu$ satisfies ${\beta^\mu}_\lambda{\beta^\lambda}_\nu={\gamma^\mu}_\nu$. Defining  
\begin{equation}\label{1f}
\langle\kappa^n\rangle\equiv g_{\mu\nu}{\kappa^\mu}_{\alpha_1}{\kappa^{\alpha_1}}_{\alpha_2}\cdots\kappa^{\alpha_{n-1}\nu}
\end{equation}
we have
\begin{eqnarray}
{\mathcal L}^{(2)}&=&\langle\kappa\rangle^2-\langle\kappa^2\rangle,\label{2a}\\
{\mathcal L}^{(3)}&=&\langle\kappa\rangle^3-3\langle\kappa\rangle\langle\kappa^2\rangle+2\langle\kappa^3\rangle,\label{2b}\\
{\mathcal L}^{(4)}&=&\langle\kappa\rangle^4-6\langle\kappa\rangle^2\langle\kappa^2\rangle+3\langle\kappa^2\rangle^2\nonumber\\&&+8\langle\kappa\rangle\langle\kappa^3\rangle
-6\langle\kappa^4\rangle\label{2c}.
\end{eqnarray}
The equations of motion associated with the above Lagrangian read
\begin{equation}\label{1g}
G^{\mu\nu}=-\frac{m^2}{\sqrt{-g}}\frac{\delta(a_i{\sqrt{-g}}{\mathcal L}^{(i)})}{\delta g_{\mu\nu}}
\end{equation} 
where $i=2,3,4$. On the other hand, from Eqs. (\ref{2a})-(\ref{2c}) we have the following relations  
\begin{eqnarray}
\frac{\delta{\mathcal L}^{(2)}}{\delta g_{\mu\nu}}&=&\kappa^{\mu\nu}-H^{\mu\nu}+\langle\kappa\rangle(g^{\mu\nu}-\kappa^{\mu\nu}),\label{3a}\\
\frac{\delta{\mathcal L}^{(3)}}{\delta g_{\mu\nu}}&=&\frac{3}{2}{\kappa^\mu}_\alpha H^{\alpha\nu}+\frac{3}{2}
{H^\mu}_\alpha\kappa^{\alpha\nu}+3H^{\mu\nu}-6\kappa^{\mu\nu}\nonumber\\&&+\frac{3}{2}{\mathcal{L}}^{(2)}
(g^{\mu\nu}-\kappa^{\mu\nu})-3\langle\kappa\rangle(H^{\mu\nu}-\kappa^{\mu\nu}),\label{3b}\\
\frac{\delta{\mathcal L}^{(4)}}{\delta g_{\mu\nu}}&=&12{H^\mu}_\alpha H^{\alpha\nu}-24H^{\mu\nu}+48\kappa^{\mu\nu}\nonumber\\&&-18{\kappa^\mu}_\alpha 
H^{\alpha\nu}-18{H^\mu}_\alpha\kappa^{\alpha\nu}\nonumber\\&&+2{\mathcal{L}}^{(3)}
(g^{\mu\nu}-\kappa^{\mu\nu})-6{\mathcal{L}}^{(2)}(H^{\mu\nu}-\kappa^{\mu\nu})\nonumber
\\&&+2\langle\kappa\rangle\left(3{\kappa^\mu}_\alpha H^{\alpha\nu}+3{H^\mu}_\alpha\kappa^{\alpha\nu}\right.\nonumber\\&&\left.+6H^{\mu\nu}-12\kappa^{\mu\nu}\right)\label{3c}
\end{eqnarray}
where ${H^\mu}_\nu=\delta^\mu_\nu-{\gamma^\mu}_\nu$. By inserting these into Eq. (\ref{1g}), the explicit form of the equations of motion are obtained. The resulting expressions are
algebraically complicated and not illuminating. In the next section we present the relevant relations for the case of a plane gravitational wave.
\section{Gravitational wave solution}
We consider a plane gravitational wave ansatz given by the following line element 
\begin{equation}\label{4a}
ds^2=-dudv-F(u,x,y)du^2+dx^2+dy^2
\end{equation}
in which $u=t-z$ and $v=t+z$ are the null coordinates and $F(u,x,y)$ is to be obtained from the equations of motion. By taking $F(u,x,y)$ equal to the product of arbitrary functions of 
$u$ and solutions of the Laplace equation (in terms of $x$ and $y$), this is an exact gravitational wave solution in general relativity. In this chart the Minkowski metric is given by
\begin{equation}\label{4bb}
\gamma_{\mu\nu} =
\left( \begin{array}{cccc}
0 & -\frac{1}{2} &0 & 0 \\ 
-\frac{1}{2} & 0 & 0 & 0 \\
0 & 0 & 1 & 0\\
0 & 0 & 0 & 1
\end{array} \right)
\end{equation}
 and hence,
\begin{equation}\label{4b}
{\gamma^\mu}_\nu =
\left( \begin{array}{cccc}
1 & 0 &0 & 0 \\ 
-2F & 1 & 0 & 0 \\
0 & 0 & 1 & 0\\
0 & 0 & 0 & 1
\end{array} \right)
\end{equation}
from which we obtain
\begin{equation}\label{4c}
{\beta^\mu}_\nu =
\left( \begin{array}{cccc}
1 & 0 &0 & 0 \\ 
-F & 1 & 0 & 0 \\
0 & 0 & 1 & 0\\
0 & 0 & 0 & 1
\end{array} \right)
\end{equation} 
There are several other square roots, but they are incompatible with the above ansatz. We thus have 
\begin{equation}\label{4d}
{\kappa^\mu}_\nu =
\left( \begin{array}{cccc}
0 & 0 &0 & 0 \\ 
F & 0 & 0 & 0 \\
0 & 0 & 0 & 0\\
0 & 0 & 0 & 0
\end{array} \right).
\end{equation} 
This results in 
\begin{equation}\label{4e}
{\mathcal{L}}^{(i)}=0,
\end{equation}
\begin{eqnarray}
\frac{\delta{\mathcal{L}}^{(3)}}{\delta g_{\mu\nu}}&=&0,\label{4f}\\
\frac{\delta{\mathcal{L}}^{(4)}}{\delta g_{\mu\nu}}&=&0,\label{4ff}
\end{eqnarray}
and  
\begin{equation}\label{4g}
\frac{\delta{\mathcal{L}}^{(2)}}{\delta g_{\mu\nu}}=
\left( \begin{array}{cccc}
0 & 0 &0 & 0 \\ 
0 & 2F & 0 & 0 \\
0 & 0 & 0 & 0\\
0 & 0 & 0 & 0
\end{array} \right).
\end{equation}
Inserting these into the field equation (\ref{1g}) and noting that for the metric given in Eq. (\ref{4a}) we have
\begin{equation}\label{4gg}
G^{\mu\nu}=
\left( \begin{array}{cccc}
0 & 0 &0 & 0 \\ 
0 & -2\nabla^2 F(u,x,y) & 0 & 0 \\
0 & 0 & 0 & 0\\
0 & 0 & 0 & 0
\end{array} \right)
\end{equation}
where $\nabla^2$ is the Laplacian in the transverse plane, we obtain
\begin{equation}\label{y6}
\nabla^2 F(u,x,y)=a_2m^2 F(u,x,y).
\end{equation}
Thus we obtain a plane wave solution provided the waveform satisfies a two-dimensional Helmholtz or modified Helmholtz equation, depending on the sign of $a_2$. Therefore  we obtain 
\begin{equation}\label{4j}
F(u,x,y)=f_1(u)F_1(x,y)+f_2(u)F_2(x,y)
\end{equation}
in which $f_1(u)$ and $f_2(u)$ are arbitrary functions of $u$ and $F_1(x,y)$ and $F_2(x,y)$ are solutions to the two dimensional (modified) Helmholtz equation. This describes a plane wave 
propagating in the $z$-direction with the speed of light. The solutions to the (modified) Helmholtz equation can be expressed, in polar coordinates, in terms of the (modified) Bessel functions. 
The singularity at $x=y=0$ then suggests a source located there. Thus the above plane wave corresponds to the gravitational wave generated by that source. In particular, this may be compared 
with the Aichelburg-Sexl solution \cite{sexl}
\begin{equation}\label{a71}
F(u,x,y)=C\delta(u)\ln(x^2+y^2)
\end{equation} 
with $C$ being a constant proportional to the particle momentum. This describes the general relativistic shock wave due to a massless point-like particle propagating longitudinally.

The modified Helmholtz equation is usually associated with diffusion-like phenomena and hence the non-negative values of $a_2$ are also disregarded from this point of view.

One can study the physical properties of the plane wave solution Eqs. (\ref{4a}) and (\ref{4j}),  by considering its effect on the motion of test particles. This can be achieved by solving 
the geodesics of the space-time for the trajectory of test particles or by taking the geodesic deviation equations into account. This has been considered for the case of plane waves of type given 
by Eq. (\ref{a71}) in several works including Refs. \cite{dray,fer,stein}. Since the equation for geodesics involves the first derivatives of $F(u,x,y)$ with respect to the transverse coordinates 
(through the connection coefficients), the effect of the graviton mass, the right-hand side of  Eq. (\ref{y6}),  can then be traced from the terms containing such derivatives.

Now it is interesting to ask if the above plane wave, Eq. (\ref{4a}),  is a solution to massive gravities with a general  nonlinear potentials given by
\begin{equation}\label{14}
U_{gen}=-{\mathcal L}^{(2)}+\sum_{i=3}{\mathcal L}^{(i)}_{gen}
\end{equation} 
in which
\begin{eqnarray}
{\mathcal L}^{(3)}_{gen}&=&C^{(3)}_1\langle{H}\rangle^3+C^{(3)}_2\langle{H}\rangle\langle{H}^2\rangle+C^{(3)}_3\langle{H}^3\rangle,\label{14a}\\
{\mathcal L}^{(4)}_{gen}&=&C^{(4)}_1\langle{H}\rangle^4+C^{(4)}_2\langle{H}\rangle^2\langle{H}^2\rangle+C^{(4)}_3\langle{H}^2\rangle^2\nonumber\\&&
+C^{(4)}_4\langle{H}\rangle\langle{H}^3\rangle+C^{(4)}_5\langle{H}^4\rangle\label{14b},\\
{\mathcal L}^{(5)}_{gen}&=&C^{(5)}_1\langle{H}\rangle^5+C^{(5)}_2\langle{H}\rangle^2\langle{H}^3\rangle+C^{(5)}_3\langle{H}\rangle^3\langle{H}^2\rangle
\nonumber\\&&+C^{(5)}_4\langle{H}\rangle\langle{H}^4\rangle+C^{(5)}_5\langle{H}^2\rangle\langle{H}^3\rangle\nonumber\\&&
+C^{(5)}_6\langle{H}\rangle\langle{H}^2\rangle^2+C^{(5)}_7\langle{H}^5\rangle\label{14c}
\end{eqnarray}
and similar expressions for higher order terms. Here $\langle H^n\rangle$ is the trace of $({H^{\mu\nu}})^n$, and $C^{(n)}_j$ are generic coefficients. Now by using the ansatz given by Eq. 
(\ref{4a}), we can repeat the previous calculations to obtain
\begin{eqnarray}
{\mathcal L}^{(n)}_{gen}&=&0,\label{15a}\\
\frac{\delta{\mathcal{L}}^{(i)}_{gen}}{\delta g_{\mu\nu}}&=&0,\label{15b}
\end{eqnarray}  
Thus the plane wave is a solution to Eq. (\ref{1g}) for massive gravities with a nonlinear potential with non-tuned coefficients. However, such models are not free of ghosts in general.  
\section{Conclusions}
We have shown that the dRGT model of massive gravity admits plane gravitational wave solutions. The waveform only depends on the quadratic term in the interaction potential and obeys the 
Helmholtz equation for negative values of $a_2$. It reduces to the general relativistic result, the Laplace equation,  for a vanishing $a_2$. Thus we obtained the generalization of the 
Aichelburg-Sexl solution in the framework of the dRGT massive gravity model. The graviton mass enters the solution via the massive Helmholtz equation. The solution is valid regardless of  
values of the coefficients in front of the cubic and quartic terms in the lagrangian. In the massless limit the solution is the same as in general relativity. The plane wave solution may be used to 
study the scattering of particles in the gravitational field of a massless particle in the context of massive theories of gravity.           
\section*{Acknowledgments} I would like to thank Gregory Gabadadze for valuable comments. This work
was supported by Payame Noor University.

\end{document}